%% file: main.tex
\documentclass{article}

\usepackage{microtype}
\usepackage{graphicx}
\usepackage{booktabs}
\usepackage{hyperref}

\usepackage[accepted]{icml2026}

\usepackage{amsmath,amssymb}
\usepackage{algorithm}
\usepackage{algorithmic}
\usepackage{multirow}
\usepackage{listings}
\usepackage{listings-turtle-paper}
\usepackage{xcolor}
\usepackage{tikz}
\usetikzlibrary{positioning,arrows.meta,shapes.geometric,automata,calc}
\usepackage{acro}

\DeclareAcronym{CFSS}{short=CFSS, long=customer-facing service specification}
\DeclareAcronym{RFSS}{short=RFSS, long=resource-facing service specification}
\DeclareAcronym{TIO}{short=TIO, long=TMF Intent Ontology}
\DeclareAcronym{SID}{short=SID, long=Shared Information and Data}
\DeclareAcronym{SHACL}{short=SHACL, long=Shapes Constraint Language}
\DeclareAcronym{LLM}{short=LLM, long=large language model}

\definecolor{kw}{RGB}{0,80,160}
\definecolor{uri}{RGB}{120,60,0}
\definecolor{str}{RGB}{0,120,0}
\definecolor{cmt}{RGB}{120,120,120}

\definecolor{color0}{RGB}{0,0,0}
\definecolor{color1}{RGB}{255,0,0}
\definecolor{color2}{RGB}{0,0,255}
\definecolor{color3}{RGB}{0,127,0}

\lstdefinestyle{turtle}{
  basicstyle=\footnotesize\ttfamily,
  keywordstyle=\color{kw}\bfseries,
  stringstyle=\color{str},
  commentstyle=\color{cmt}\itshape,
  morekeywords={a},
  morestring=[b]",
  morecomment=[l]{\#},
  columns=fullflexible,
  keepspaces=true,
  showstringspaces=false,
  frame=none,
  aboveskip=4pt,
  belowskip=2pt,
  xleftmargin=4pt,
  xrightmargin=2pt,
  breaklines=true,
  captionpos=b,
  numberstyle=\tiny\color{cmt},
  escapeinside={(*@}{@*)},
}

\newcommand{\RETURN}{\STATE \textbf{return} }

\icmltitlerunning{Intent-Driven 6G Service Orchestration: Grounded Translation, Validation, and Decomposition}
\begin{document}

\twocolumn[

\icmltitle{Intent-Driven 6G Service Orchestration: Grounded Translation, \\Validation, and Decomposition}

\begin{icmlauthorlist}
\icmlauthor{Jean Martins}{er}
\icmlauthor{Leonid Mokrushin}{er}
\icmlauthor{Marin Orlic}{er}
\icmlauthor{Amardeep Kumar A}{bcss}
\end{icmlauthorlist}

\icmlaffiliation{er}{Ericsson Research}
\icmlaffiliation{bcss}{Ericsson BCSS BOS PDU}
\icmlcorrespondingauthor{Jean Martins}{jean.martins@ericsson.com}

\icmlkeywords{Intent-Based Networking, Agentic AI, LLM, 6G, Constraint Satisfaction, SHACL, Semantic Catalog}

\vskip 0.3in
]

\printAffiliationsAndNotice{}

\begin{abstract}
Intent-based automation for 6G envisions networks steered by high-level goals rather than low-level configurations. Existing LLM-based approaches translate natural language into plausible intent representations but typically omit what production deployment requires: grounding in actual service catalogs, formal validation, and cross-layer decomposition. We address this with an agentic workflow comprising three coupled reasoning layers: (i) \emph{grounding} the translation in a semantic service catalog that exposes TMF-compliant service  specifications; (ii) \emph{validation} of the RDF intent via SHACL structural checking against the TMF Intent Ontology; and (iii) \emph{decomposition} that selects a CFSS profile via constraint satisfaction over QoS capability envelopes, then covers its infrastructure requirements with RFSS profiles via weighted set cover. Across 930 benchmark runs over six GPT-4.1/5 models, the workflow achieves 97\% success in structured mode and 90\% on average across natural-language scenarios, with 100\% correct rejection of infeasible requests. Grounding LLM context in catalog capability metadata reduces adversarial hallucinations by 26 percentage points; larger gains than scaling model size alone.
\end{abstract}

\input{sections/01-introduction}
\input{sections/015-background}
\input{sections/02-model}
\input{sections/03-architecture}

\input{sections/04-experiments}
\input{sections/05-related}

\input{sections/06-conclusion}

\bibliography{references}
\bibliographystyle{icml2026}


\newpage
\appendix
\input{sections/appendix}

\end{document}

%% file: sections/01-introduction.tex
\section{Introduction}
\label{sec:introduction}

The evolution toward 6G demands intent-driven service orchestration: applications and operators express high-level service requirements and the management system validates, decomposes, and fulfills them as formal service orders~\citep{rfc9315, ibnsurvey}. The TM~Forum defines two complementary frameworks for this pipeline: the \ac{SID} model,\footnote{\url{https://www.tmforum.org/open-digital-architecture/information-framework-sid/}} which structures service catalogs into \acp{CFSS} and \acp{RFSS} with explicit QoS capability envelopes, and the Intent Common Model~\citep{tmf-tr290}, which standardizes intent expression as RDF graphs over the \ac{TIO}, aligned with ETSI ZSM~\citep{etsi-zsm-002} and 3GPP management frameworks. Bridging these two, turning an intent into a validated service order against the \ac{SID} catalog; aligns with the foreseen evolution toward intent-driven autonomous networks~\citep{ericsson-intent-evolution}, yet remains largely unsolved.

As \citet{ficzere2025beyond} argue, current work on Generative~AI for IBN remains narrowly focused on intent \emph{translation}, leaving substantial gaps across the rest of the IBN life cycle, notably feasibility validation, orchestration composition, and structured reporting. Recent \ac{LLM}-based approaches to intent and slice management~\citep{mekrache2025dmo, ibn-llm-centric, ibn-autonomous} generate intent expressions from natural language, but typically skip three capabilities that are essential for telecom service management~\citep{ericsson-intent-evolution}: \emph{structural validation} of the generated intent against the \ac{TIO}, \emph{feasibility reasoning} against the QoS capability envelopes declared in the \ac{SID} catalog, and \emph{cross-layer decomposition} of customer-facing selections into resource-facing infrastructure (network slice, edge compute, transport). Without all three, an \ac{LLM} can produce syntactically plausible but semantically invalid or infeasible service orders that no intent handler should accept.

To illustrate, consider an operator requesting ``cloud gaming with latency $\leq$ 10\,ms.'' Under this model, the intent triggers a service order~\citep{tmf641} against the service catalog~\citep{tmf633}: the system should (1)~generate a structurally valid RDF intent conforming to \ac{TIO}; (2)~discover that the Pro~Gaming \ac{CFSS} (latency envelope 5--10\,ms) satisfies the constraint while Casual~Gaming (50--100\,ms) does not; and (3)~decompose Pro~Gaming's infrastructure requirements into \acp{RFSS}: a URLLC network-slice profile, an edge-compute pool, and a transport segment. This pipeline goes well beyond intent translation: it requires grounding, validation, and decomposition as first-class concerns.

We present an agentic system that addresses this end-to-end. It sits at the natural-language boundary of a TMF intent-management hierarchy~\citep{tmf-ig1253}, produces validated RDF intents, and emits a \ac{CFSS} selection together with an \ac{RFSS} decomposition that downstream intent handlers can consume. The system does \emph{not} attempt to configure RAN or transport equipment directly; instead, it grounds \ac{LLM} generation in a semantic service catalog and a formal RequirementCapability model, so that ``what'' is requested and ``what can be delivered'' are explicitly reconciled before a plan leaves the NL layer.


%% file: sections/015-background.tex
\section{Background}
\label{sec:background}

\subsection{TMF Intent Ontology}

The \ac{TIO}~\citep{tmf-tr290} is the RDF vocabulary defined for expressing network management intents. It is specified across 15 ontology modules covering intent structure, logical composition, quantity constraints, metrics, and lifecycle management. The core constructs include \texttt{icm:Intent} as the top-level container, expectations for requirements (specialized into delivery, property, and reporting), and targets for scope. Logical operators (\texttt{allOf}, \texttt{anyOf}) compose expectations into boolean expressions, and the Quantity Ontology provides metric-based constraints (\texttt{atLeast}, \texttt{atMost}) with unit handling for properties such as latency. Listing~\ref{lst:intent-example} shows a minimal intent requesting cloud gaming with latency $\leq$ 10\,ms. 

\begin{figure}[h]
\begin{lstlisting}[language=TurtlePaper,numbers=none,basicstyle=\ttfamily\scriptsize,caption={\ac{TIO} intent: cloud gaming with latency $\leq$ 10\,ms.},label=lst:intent-example]
@prefix icm:  <http://tio.../IntentCommonModel/> .
@prefix log:  <http://tio.../LogicalOperators/> .
@prefix met:  <http://tio.../MetricsAndObservations/> .
@prefix quan: <http://tio.../QuantityOntology/> .
@prefix dim:  <http://tio.../Dimensions/> .

ex:GamingIntent a icm:Intent ;
    log:allOf ( ex:DeliveryExp ex:LatencyExp ) .

ex:DeliveryExp a icm:DeliveryExpectation ;
    icm:target [ a icm:Target ] ;
    icm:deliveryType cfss:GamingCFSS .

ex:LatencyCond a log:Condition ;
    quan:atMost ( [ met:lastValue ( dim:Latency ) ]
                  "10ms"^^quan:quantity ) .

ex:LatencyExp a icm:PropertyExpectation ;
    log:allOf ( ex:LatencyCond ) .
\end{lstlisting}
\end{figure}

Composing such intents correctly is error-prone due to \ac{TIO}'s recursive logical operators, quantity expressions, and cross-expectation dependencies; guaranteeing ontology compliance before an intent reaches a handler is therefore essential. We leverage TIO-SHACL~\citep{martins2026tioshacl} to provide structural validation against the ontology.

\subsection{SID Service Catalog}

The TM~Forum Information Framework (\ac{SID}) is a comprehensive reference model that organizes telecom concepts across product, service, and resource domains. In this work we focus on the \emph{service domain} used by service catalogs: \acp{CFSS} define application-facing service profiles with QoS capability envelopes visible to customers and operators (e.g., as exposed through TMF~633/641), while \acp{RFSS} describe the underlying infrastructure capabilities (network slices, compute pools, transport segments) required to deliver those services. A \ac{CFSS} profile typically requires several \ac{RFSS} profiles working together; resolving this many-to-many mapping between layers is precisely the decomposition problem we address. In ODA terms, our system sits between the intent interface and TMF~633/641-driven service order management.

%% file: sections/02-model.tex
\section{Semantic Service Catalog and Problem Formulation}
\label{sec:model}

\begin{figure}[h]
\centering
\begin{tikzpicture}[
  >=Stealth,
  scale=0.85, every node/.style={transform shape},
  abstract/.style={rectangle, draw=black, fill=black!8, thick, dashed, minimum width=1.9cm, minimum height=0.38cm, font=\scriptsize\itshape},
  type/.style={rectangle, draw=black, fill=color3!15, thick, minimum width=1.9cm, minimum height=0.38cm, font=\scriptsize},
  profile/.style={rectangle, draw=black, fill=color3!8, thick, minimum width=1.8cm, minimum height=0.34cm, font=\scriptsize},
  rfss_type/.style={rectangle, draw=black, fill=color1!15, thick, minimum width=1.9cm, minimum height=0.38cm, font=\scriptsize},
  rfss_profile/.style={rectangle, draw=black, fill=color1!8, thick, minimum width=1.8cm, minimum height=0.34cm, font=\scriptsize}
]
  \node[abstract] (cfss_parent) {CFSSParent};
  \node[type, below=0.5cm of cfss_parent] (gaming) {GamingCFSS};
  \node[profile, below left=0.5cm and 0.8cm of gaming] (casual) {Casual};
  \node[profile, below=0.5cm of gaming] (competitive) {Competitive};
  \node[profile, below right=0.5cm and 0.8cm of gaming] (pro) {Pro};

  \draw[->, thick, dashed] (cfss_parent) -- (gaming);
  \draw[->, thick, dashed] (gaming) to[out=-135, in=90] (casual);
  \draw[->, thick, dashed] (gaming) -- (competitive);
  \draw[->, thick, dashed] (gaming) to[out=-45, in=90] (pro);

  \node[rfss_profile, below=1.6cm of casual] (urllc) {URLLC Premium};
  \node[rfss_profile, below=1.6cm of competitive] (edge) {Premium Edge};
  \node[rfss_profile, below=1.6cm of pro] (transport) {Prem.\ Transport};
  \node[rfss_type, below=0.5cm of edge] (slice) {NetworkSliceRFSS};
  \node[abstract, below=0.5cm of slice] (rfss_parent) {RFSSParent};

  \draw[->, thick, dashed] (slice) to[out=135, in=-90] (urllc);
  \draw[->, thick, dashed] (slice) -- (edge);
  \draw[->, thick, dashed] (slice) to[out=45, in=-90] (transport);
  \draw[->, thick, dashed] (rfss_parent) -- (slice);

  \draw[->, thick, color=color2] (pro) -- node[right, font=\scriptsize, color=color2] {\emph{sat.}} (urllc);
\end{tikzpicture}
\caption{Semantic service catalog hierarchy. CFSS profiles (green) declare QoS capability envelopes; RFSS profiles (red) provide infrastructure. The blue arrow denotes a satisfaction relationship computed by the matching algorithm.}
\label{fig:catalog-hierarchy}
\end{figure}
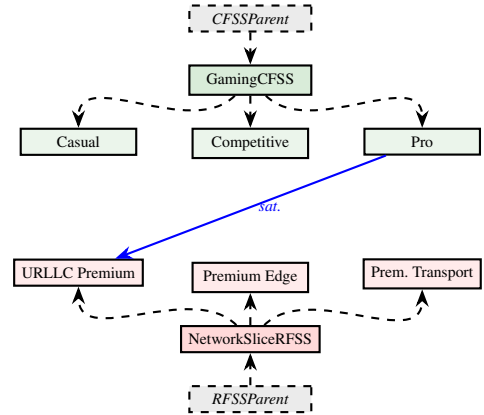

To enable formal reasoning over the \ac{SID} service domain, we build a semantic layer that makes QoS capabilities and requirements explicit and machine-actionable. This layer serves as the grounding infrastructure for the agentic pipeline: it provides the formal substrate against which an \ac{LLM}-generated intent can be validated for feasibility, matched to a \ac{CFSS} profile, and decomposed into \ac{RFSS} selections. Figure~\ref{fig:catalog-hierarchy} illustrates the resulting catalog hierarchy, where CFSS profiles declare QoS capability envelopes and RFSS profiles provide the underlying infrastructure. 

\subsection{Requirement-Capability ontology}

The RequirementCapability model introduces two dual predicates over this hierarchy (Figure~\ref{fig:catalog-hierarchy}). The \emph{provides} predicate declares what a profile can deliver for a given metric (a range with minimum, maximum, unit, or an enumeration such as URLLC/eMBB). The \emph{requires} predicate declares what an intent or CFSS profile demands from the layer below. Each layer acts as both consumer and provider: an intent requires CFSS capabilities; a CFSS profile provides capabilities to intents and requires RFSS capabilities; RFSS profiles provide capabilities to CFSS profiles.

A \emph{capability} $c \in C(p)$ for profile $p$ specifies a metric with a deliverable range $[c^-, c^+]$ and unit, or an enumeration of allowed values. A \emph{requirement} is defined analogously. Appendix~\ref{app:ontology} shows the RDF serialization of these predicates.

\subsection{Problem Formulation}

Given a natural-language intent and a service catalog, the intent-driven service management problem decomposes into four sub-problems. Structural correctness of the generated RDF intent (\ac{SHACL} validation against \ac{TIO}) is a prerequisite enforced at generation time; the sub-problems below concern the end-to-end pipeline from natural language to validated service order.

\textbf{P0 (Intent Translation).} Generate a structurally valid \ac{TIO}/RDF intent from a natural-language requirement, passing \ac{SHACL} validation.

\textbf{P1 (Constraint Extraction).} Extract from the validated intent a structured set of QoS requirements $R = \{r_1, \ldots, r_n\}$ where each $r_i = (m_i, \mathit{op}_i, v_i, u_i)$ is a tuple of metric, operator, value and unit (or $(m_i, \mathit{op}_i, e_i)$ for enumerations). 

\textbf{P2 (CFSS Selection).} Given requirements $R$ and CFSS profiles $P = \{p_1, \ldots, p_m\}$ with capabilities $C(p_j)$, find a minimum-cost profile satisfying all requirements:
\begin{equation}
p^* = \arg\min_{p \in P}\; \mathit{cost}(p) \;\;\text{s.t.}\;\; \forall\, r \in R,\; \exists\, c \in C(p)\!: \operatorname{Sat}(c, r).
\label{eq:cfss-matching}
\end{equation}

\textbf{P3 (RFSS Decomposition).} Given the selected CFSS profile's infrastructure requirements $R_{\text{cfss}}$ and candidate RFSS profiles $Q$ (network-slice, edge, transport) with capabilities $C(q_i)$ and costs $w(q_i)$, find the minimum-cost subset covering all requirements:
\begin{equation}
S^* = \arg\min_{S \subseteq Q}\; {\textstyle\sum_{q \in S}} w(q) \;\;\text{s.t.}\;\; {\textstyle\bigcup_{q \in S}} C(q) \supseteq R_{\text{cfss}}.
\label{eq:rfss-decomp}
\end{equation}

P2 is a capability-set constraint-satisfaction problem over profile choice; P3 is weighted set cover over infrastructure profiles, which we solve with a greedy heuristic (\mbox{$O(\ln n)$} approximation) and compare against a CP-SAT solver.

\subsection{Two-Layer Validation}

The system combines two validation layers. \emph{\ac{SHACL} shapes}~\citep{martins2026tioshacl} validate structural correctness of the intent RDF against \ac{TIO}: required fields, correct types, and cardinality. The \emph{RequirementCapability} layer provides feasibility reasoning: range- and enumeration-based predicates over capability sets determine whether any catalog profile can satisfy the extracted requirements. SHACL catches missing or malformed constructs (e.g., a missing metric field), while RequirementCapability catches semantic infeasibility (e.g., requesting a QoS level no profile can deliver). This separation enables the system to distinguish fixable structural errors from fundamentally infeasible requirements.

%% file: sections/03-architecture.tex
\section{Agentic Architecture}
\label{sec:architecture}

The system is implemented as a LangGraph\footnote{\url{https://github.com/langchain-ai/langgraph}} state machine that orchestrates LLM calls and external tool invocations. Rather than a RAN or transport control-plane component, the system sits at the NL boundary of a TMF-style intent-management hierarchy: it ingests a natural-language requirement, grounds it in the semantic service catalog, emits a validated TIO/RDF intent together with a CFSS selection and an RFSS decomposition plan, and hands the result to downstream intent handlers for execution. 3GPP RAN and transport controllers remain out of scope.

\subsection{Dual-Mode Workflow}

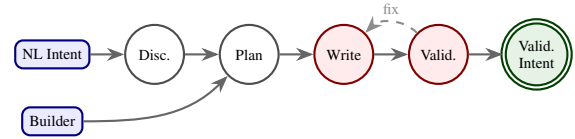
\begin{figure}[h]
\centering
\begin{tikzpicture}[
  >=Stealth,
  node distance=0.55cm,
  every node/.style={transform shape},
  scale=0.82,
  state/.style={circle, draw=black!70, thick, minimum size=0.95cm, font=\scriptsize, align=center, fill=white},
  entry/.style={rectangle, rounded corners=3pt, draw=color2!50!black, thick, minimum width=0.9cm, minimum height=0.45cm, font=\scriptsize, fill=color2!8},
  shared/.style={state, fill=color1!8, draw=color1!50!black},
  output/.style={state, fill=color3!10, draw=color3!50!black, double, double distance=0.8pt},
  arr/.style={->, thick, black!60},
  fix/.style={->, thick, dashed, black!45}
]
  \node[entry] (nl) {NL Intent};
  \node[entry, below=0.6cm of nl] (ui) {Builder};
  \node[state, right=of nl] (disc) {Disc.};
  \node[state, right=of disc] (plan) {Plan};
  \node[shared, right=of plan] (write) {Write};
  \node[shared, right=of write] (validate) {Valid.};
  \node[output, right=of validate] (done) {Valid.\\Intent};
  \draw[arr] (nl) -- (disc);
  \draw[arr] (disc) -- (plan);
  \draw[arr] (ui) to[out=0, in=-135] (plan);
  \draw[arr] (plan) -- (write);
  \draw[arr] (write) -- (validate);
  \draw[arr] (validate) -- (done);
  \draw[fix] (validate) to[out=135, in=45] node[above, font=\scriptsize] {fix} (write);
\end{tikzpicture}
\caption{Dual-mode agentic workflow. NL mode includes catalog discovery; Builder mode bypasses directly to planning. Both converge at generation and validation with a self-healing loop (dashed).}
\label{fig:workflow}
\end{figure}

The system supports two entry modes (Figure~\ref{fig:workflow}). \textbf{NL mode} processes free-text requirements through discovery (semantic search over the catalog to identify candidate CFSS families), capability retrieval, and plan decomposition. \textbf{Builder mode} accepts structured inputs specifying the target CFSS profile and requirement set directly, bypassing NL processing for programmatic integration with reduced token consumption. Both modes converge on the same plan structure that drives intent generation and validation.

\subsection{Intent Generation and Self-Healing}

The write-intent stage transforms the plan into RDF/Turtle. The prompt combines a stable prefix containing TIO syntax rules, conditionally injected validated patterns retrieved by fuzzy matching, and a dynamic context comprising the requirement, available-property metadata from the catalog, and the plan. Available-property metadata extracted from the catalog's \texttt{req:provides} blocks is injected explicitly so that the model is told which QoS metrics the target service models (see Section~\ref{sec:experiments}).

After generation, the RDF passes through a two-stage validation pipeline with automatic error correction. Algorithm~\ref{alg:pipeline} shows the end-to-end composition: LLM-based NL-to-RDF translation (P1), SHACL structural validation, CSP-style CFSS selection (P2), and weighted-set-cover RFSS decomposition (P3) are composed as a single pipeline, with a bounded retry combinator $\textsc{Retry}_k$ providing self-healing around the fallible LLM step. The routing logic distinguishes three outcomes: both checks pass (accept the intent); SHACL fails (retry generation with violation context); feasibility fails (terminate---the intent is structurally valid but cannot be satisfied by any catalog profile).

\begin{algorithm}[t]
\caption{Grounded intent composition. Components defined in Section~\ref{sec:model}.}
\label{alg:pipeline}
\begin{algorithmic}[1]
\REQUIRE Requirement $\rho$ (NL or structured), catalog $(P, Q)$, shapes $\Sigma$
\ENSURE Validated intent $I$ with CFSS $p^*$ and RFSS cover $S^*$, or $\bot$
\STATE $\mathcal{M} \gets \textsc{Ground}(\rho, P)$ \COMMENT{available-property metadata}
\STATE $I \gets \textsc{Retry}_k\bigl[\textsc{LLMWrite}(\rho, \mathcal{M}, \Sigma)\ \textsc{st}\ \textsc{Shacl}(I, \Sigma)\bigr]$
\IF{$I = \bot$} \RETURN $\bot$ \COMMENT{structural repair failed} \ENDIF
\STATE $R \gets \textsc{Extract}(I)$ \COMMENT{P1: constraint set}
\STATE $p^* \gets \textsc{SatCfss}(R, P)$ \COMMENT{P2: CSP, Eq.~\ref{eq:cfss-matching}}
\IF{$p^* = \bot$} \RETURN $\bot$ \COMMENT{infeasible: no CFSS covers $R$} \ENDIF
\STATE $S^* \gets \textsc{CoverRfss}(R_{\text{cfss}}(p^*), Q)$ \COMMENT{P3: Eq.~\ref{eq:rfss-decomp}}
\RETURN $(I, p^*, S^*)$
\end{algorithmic}
\end{algorithm}

\textsc{Retry}$_k$ invokes \textsc{LLMWrite} up to $k$ times (we use $k=5$ for both syntactic and semantic correction), each retry seeded with the prior SHACL violation report and with validated patterns retrieved via fuzzy matching. This mechanism is critical for deployability: rather than silently producing an invalid or infeasible intent, the system either converges to a valid one or explicitly terminates with a diagnosis (line~3 for unrepairable SHACL violations, line~6 for CFSS-level infeasibility).

%% file: sections/04-experiments.tex
\section{Experimental Evaluation}
\label{sec:experiments}

We evaluate the system across six LLM models using five benchmark configurations totaling 930 runs. The evaluation measures end-to-end success rates, per-phase pipeline accuracy, and the impact of prompt difficulty and context grounding on model performance.

\subsection{Setup}

The six models span two Azure OpenAI families, GPT-4.1 (nano, mini, full) and GPT-5 (nano, mini, full), each executed five times per input at temperature 0.0. The catalog contains five CFSS families (Cloud Gaming, Video Streaming, Video Conferencing, Audio, Data Connectivity) with multiple profiles per family, plus the three RFSS categories described in Section~\ref{sec:model} (network-slice, edge-compute, transport). Profiles declare QoS capability envelopes over metrics such as latency, jitter, throughput, bandwidth, packet loss, availability, and enumeration metrics such as slice type (URLLC/eMBB/mMTC) and 3GPP 5QI.

The benchmark configurations form two axes. The first axis is \emph{input mode}: NL (free-text requirement, full pipeline) vs.\ Builder (structured input, direct plan). The second axis is \emph{difficulty}: constrained, adversarial, infeasible.

\begin{table}[h]
\centering
\scriptsize
\begin{tabular}{@{}llrl@{}}
\toprule
Config & Mode & Runs & Description \\
\midrule
CFSS & Builder & 150 & Structured input, write+validate only \\
RFSS & Builder & 270 & Decomposition targets (slice, edge, transport) \\
Adversarial & Builder & 120 & Additional validity/reporting clauses \\
Infeasible & Builder & 90 & Impossible QoS values \\
Constrained & NL & 150 & Valid constraints on modeled metrics \\
Adversarial & NL & 120 & References to unmodeled metrics \\
Infeasible & NL & 90 & Values exceeding all profile envelopes \\
\bottomrule
\end{tabular}
\caption{Benchmark configurations (930 total runs).}
\label{tab:configs}
\end{table}

End-to-end NL evaluation targets CFSS selection, constraint extraction, and RDF generation; RFSS decomposition is evaluated through Builder~RFSS (structured) and, separately, through the CP-SAT vs.\ greedy algorithmic benchmark described in Appendix~\ref{app:catalog}.

\textbf{Success criteria.} A run is successful when the pipeline produces the expected terminal outcome: for Builder and NL~Constrained/Adversarial, a structurally valid RDF intent passing both SHACL and feasibility checks \emph{and} selecting the ground-truth profile (exact child class when the prompt names a specific profile, any class within the same CFSS family when the prompt uses only numeric constraints); for NL~Infeasible, explicit rejection with an infeasibility diagnosis ($\bot$ in Algorithm~\ref{alg:pipeline}). Silently producing a valid-looking but incorrect intent counts as a failure in all scenarios.

\subsection{End-to-End Results}

Table~\ref{tab:success-rates} presents per-model success rates. Builder mode is strong overall (97\% CFSS / 100\% RFSS), with GPT-4.1-nano's 84\% CFSS being the sole outlier, and NL mode reveals significant model-dependent 
variation, particularly under adversarial conditions.

All models achieve \textbf{100\% on infeasible intents}: the system robustly rejects QoS requests that exceed every CFSS profile's capability envelope rather than generating an invalid intent. This is essential for safe deployment, where silently accepting an infeasible intent could later propagate into SLA violations.

\begin{table}[h]
\centering
\caption{Success rates (\%) by model and experiment. Builder~C/R = Builder CFSS/RFSS; Cnst./Adv./Inf.\ = NL Constrained/Adversarial/Infeasible. Builder runs use structured inputs; NL runs exercise the full NL pipeline.}
\label{tab:success-rates}
\small
\begin{tabular}{lrrrrr}
\toprule
Model & Bld.\ & Bld.\ & NL & NL & NL \\
      & CFSS  & RFSS  & Cnst. & Adv. & Inf. \\
\midrule
GPT-4.1-nano & 84 & 100 & 52 & 60 & 100 \\
GPT-4.1-mini & 100 & 100 & 100 & 90 & 100 \\
GPT-4.1 & 100 & 100 & 100 & 90 & 100 \\
GPT-5-nano & 98 & 100 & 88 & 90 & 100 \\
GPT-5-mini & 100 & 100 & 100 & 95 & 100 \\
GPT-5 & 100 & 100 & 100 & 95 & 100 \\
\midrule
\textbf{Overall} & \textbf{97} & \textbf{100} & \textbf{90} & \textbf{87} & \textbf{100} \\
\bottomrule
\end{tabular}
\vspace{1mm}

{\footnotesize Infeasible success = correctly rejecting QoS requests beyond all CFSS capabilities.}
\end{table}

\subsection{LLM Analysis: Generation and Size}

\begin{figure}[h]
\centering
\includegraphics[width=\columnwidth]{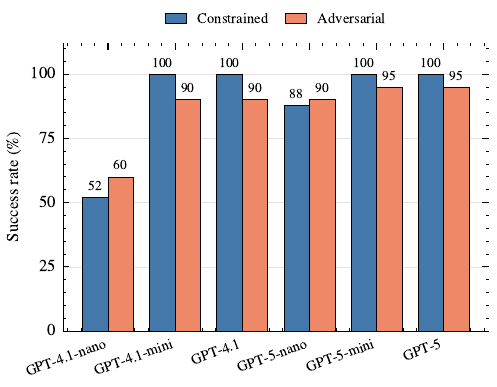}
\caption{Success rates across NL constrained and adversarial experiments. Hatched bars indicate adversarial results.}
\label{fig:success-comparison}
\end{figure}

Figure~\ref{fig:success-comparison} highlights two patterns.

\textbf{Generational leap at the nano tier.} GPT-5-nano matches GPT-4.1-full on adversarial prompts (90\% vs.\ 90\%) and closes most of the constrained-prompt gap (88\% vs.\ 100\%), a substantial jump from GPT-4.1-nano's 52--60\%. This suggests that next-generation mini- and nano-tier models may soon be viable candidates wherever intent processing needs to run on cost-constrained platforms, although our experiments use cloud-hosted models and do not themselves test edge or on-device deployment.

\textbf{Generation can rival size.} GPT-5-mini (95\% adversarial) marginally exceeds GPT-4.1-full (90\% adversarial), suggesting that within the GPT-4.1$\to$GPT-5 transition, a generational step at the mini tier competes with the prior generation's full model on structured intent tasks. We report this as observed on a single benchmark with 120 adversarial runs per model rather than as a general law.

Per-phase analysis (Appendix~\ref{app:phases}) isolates \emph{which} stage drives these numbers. Discovery and plan decomposition are essentially solved for all models ($\geq$92\%, except GPT-4.1-nano at 85--100\%), whereas \emph{constraint extraction} is the dominant bottleneck: GPT-4.1-nano reaches only 48--60\% constraint precision, while GPT-5-mini reaches 95--100\%. The end-to-end spread in Table~\ref{tab:success-rates} is almost entirely attributable to constraint-extraction error, not RDF generation or feasibility reasoning. We also observe that GPT-5 is aggressive on adversarial prompts: when a prompt mentions ``ultra-low latency'' next to the unmodeled ``frame rate,'' GPT-5 sometimes drops both rather than risk an infeasible intent, trading constraint completeness for feasibility.

\subsection{Context Grounding Reduces QoS Hallucination}

A recurring failure during development was that models invented QoS constraints for metrics not present in the catalog (e.g., creating a \texttt{FrameRate} constraint for gaming profiles). We addressed this by extracting available QoS metric names from the service ontology's \texttt{req:provides} blocks and injecting them into the LLM context as ``Available properties: Latency, Bandwidth, Jitter,\ldots'' with prompt rules telling the model to constrain only listed metrics.

Comparing otherwise-identical adversarial runs with and without this injection (Appendix~\ref{app:ablation}, Table~\ref{tab:ablation}), overall adversarial success improves from 61\% to 87\% (+26 percentage points). The effect is strongest at the mini tier, where GPT-4.1-mini rises from 50\% to 90\% and GPT-5-mini from 50\% to 95\%; larger models benefit less because they already ignore unmodeled concepts more reliably. The broader takeaway: \textbf{grounding LLM context in explicit capability metadata} yields larger reliability gains than scaling model size alone.

\subsection{Runtime and Scalability}

Since token consumption directly translates to operational cost, we compare the two modes: Builder is 40\% faster (11.7\,s vs.\ 16.4\,s mean), uses 49\% fewer tokens (4{,}582 vs.\ 6{,}848 mean), and requires one LLM call versus three. This makes Builder suitable for high-volume programmatic integration while NL mode retains flexibility for operator interaction.

Real operator catalogs are orders of magnitude larger than our benchmark; to assess scalability, we independently benchmark the P3 matching algorithms (Appendix~\ref{app:catalog}, Table~\ref{tab:catalog-perf}). The greedy heuristic runs 1.8$\times$ faster than the CP-SAT solver with identical requirement-satisfaction rates and a bounded cost gap, suggesting that online orchestration can rely on greedy at sub-second scale, with CP-SAT reserved for offline refinement as catalogs grow.

%% file: sections/05-related.tex
\section{Related Work}
\label{sec:related}

\textbf{LLM-based intent management.}
DMO-GPT~\citep{mekrache2025dmo} uses multi-agent LLMs with hierarchical planning for distributed 6G management across heterogeneous OSSs; the LLM-centric lifecycle architecture of \citet{ibn-llm-centric} demonstrates NL-driven decomposition, translation, and activation on a 5G testbed; and the IAN framework~\citep{ibn-autonomous} combines RAG-enhanced LLMs with resource allocation models. All three generate intent representations from natural language but omit formal structural validation and catalog-grounded feasibility checking. LUMI~\citep{lumi-trust} addresses trust in NL-based IBN through self-correction and verification of deployed policies, yet operates on device configurations rather than ontology-compliant RDF intents. \citet{tageldien2025llmibn} survey LLM integration across all five IBN lifecycle phases and confirm that translation dominates current research while validation and assurance remain underexplored. \citet{ficzere2025beyond} reach the same conclusion from a task-classification perspective, identifying feasibility validation and structured reporting as open problems. Our system directly addresses both gaps within a single agentic pipeline.

\textbf{Semantic and ontology-based approaches.}
SWIFT~\citep{alcock2025swift} is the closest prior work: it employs an ontology abstracting 3GPP, ETSI NFV-MANO, and ONOS data models to translate TMF-modelled slice intents into infrastructure configurations via logical programming. However, SWIFT targets imperative configuration generation rather than validated RDF intent production, and its reasoner does not perform QoS-envelope constraint satisfaction or cross-layer decomposition. \citet{mehmood2023knowledge} use knowledge graphs (OWL/RDF) to model intents extending the TMF Intent Common Model, demonstrating the value of semantic representations but without SHACL validation or LLM integration. Our work combines both directions: LLM-driven generation grounded in a semantic catalog with formal SHACL validation and CSP-based decomposition.

\textbf{6G architectures and standards.}
The 6G-INTENSE architecture~\citep{boutouchent20256gintense} proposes intent-driven management at all layers---Tenant Management Platform, Domain Manager and Orchestrator, and Network-Compute Fabric---with Native AI toolkits including LLM-powered cognitive intent handlers. Our system implements the concrete NL-to-validated-intent component that such architectures require at the tenant-to-DMO interface. \citet{brenes2025unified} present a unified architecture harmonizing 3GPP and TMF intent standards with LLM-based translation and closed-loop automation, validating the need for interoperable intent management but without formal feasibility reasoning. \citet{kapoor2025genai} survey the broader IBN+GenAI standardization landscape across ETSI, 3GPP, TMF, and industry deployments, noting that most implementations remain at Level~2--3 automation; our catalog-grounded validation and decomposition pipeline targets the formal guarantees needed for Level~4. IETF RFC~9315~\citep{rfc9315} defines intent concepts; TMF ICM~\citep{tmf-tr290} structures intents as RDF; ETSI ZSM~\citep{etsi-zsm-002} specifies closed-loop automation. None provides formal constraint satisfaction for intent-to-resource decomposition---the gap our RequirementCapability model fills.

%% file: sections/06-conclusion.tex
\section{Conclusion}
\label{sec:conclusion}

We presented an agentic LLM pipeline that goes beyond NL-to-RDF intent translation by composing three reasoning layers: catalog grounding over a semantic service specification, SHACL structural validation against the TMF Intent Ontology, and CSP-style decomposition (CFSS selection plus weighted set cover over RFSS profiles). A RequirementCapability ontology gives the three layers a shared formal substrate, so that an LLM's plausible-looking intent is only accepted when it is both well-formed and feasible against catalog capabilities. Across 930 benchmark runs spanning six GPT-4.1/5 models, the system reaches 97\% success in Builder mode and 90\% averaged across NL scenarios, with 100\% correct rejection of infeasible requests.

Three findings stand out. First, \emph{catalog-grounded context} reduces adversarial-prompt QoS hallucination by 26~percentage points, a larger gain than scaling model size alone. Second, \emph{model generation can rival parameter count}: GPT-5-mini matches or slightly exceeds GPT-4.1-full on adversarial scenarios, suggesting that next-generation mini-tier models may suffice for cost-constrained intent processing. Third, \emph{constraint extraction remains the bottleneck}: discovery, planning, and RDF generation are largely solved; the hard problem is extracting precise QoS constraints from ambiguous NL requirements.

%% file: sections/appendix.tex
\section{NL Intent Examples}
\label{app:intents}

Table~\ref{tab:nl-examples} shows representative NL intents from each benchmark configuration.

\begin{table}[ht]
\centering
\caption{Example NL intents by benchmark category.}
\label{tab:nl-examples}
\scriptsize
\begin{tabular}{@{}lp{6cm}@{}}
\toprule
Category & Example NL intent \\
\midrule
Constrained & ``I need a cloud gaming service with latency at most 10ms'' \\
 & ``I need a data service with bandwidth at least 50 Mbps and availability at least 99.5 percent'' \\
\midrule
Adversarial & ``I need a premium gaming experience with ultra-low latency and high frame rate valid during business hours with state change reporting'' \\
\midrule
Infeasible & ``I need a video conferencing service with capacity at least 5000 participants'' \\
\bottomrule
\end{tabular}
\end{table}

\section{RequirementCapability Ontology}
\label{app:ontology}

Listing~\ref{lst:provides} shows a CFSS QoS capability.

\begin{figure}[ht]
\begin{lstlisting}[language=TurtlePaper,numbers=none,basicstyle=\ttfamily\scriptsize,caption={Pro~Gaming latency capability.},label=lst:provides,aboveskip=2pt,belowskip=2pt]
cfss:ProGaming req:provides [
    a req:Capability ;
    req:metric   dim:Latency ;
    req:minValue "5"^^xsd:decimal ;
    req:maxValue "10"^^xsd:decimal ;
    req:unit     "ms" ] .
\end{lstlisting}
\end{figure}

\section{Per-Phase Pipeline Accuracy}
\label{app:phases}

Tables~\ref{tab:phases-constrained} and~\ref{tab:phases-adversarial} present per-phase accuracy for NL constrained (150 runs) and adversarial (120 runs) experiments. Discovery and planning are largely solved ($\geq$92\%); constraint extraction is the dominant bottleneck, particularly for nano-tier models.

\begin{table}[ht]
\centering
\caption{Per-phase accuracy (\%): NL constrained (150 runs).}
\label{tab:phases-constrained}
\scriptsize
\begin{tabular}{@{}lrrrrrr@{}}
\toprule
Model & Disc. & C.Prec & C.Rec & Prof. & RDF & Feas. \\
\midrule
4.1-nano & 100 & 48 & 52 & 52 & 48 & 52 \\
4.1-mini & 100 & 100 & 100 & 100 & 100 & 100 \\
4.1      & 100 & 100 & 100 & 100 & 100 & 100 \\
5-nano   & 92 & 88 & 88 & 88 & 88 & 88 \\
5-mini   & 100 & 100 & 100 & 100 & 100 & 100 \\
5        & 100 & 100 & 100 & 100 & 100 & 100 \\
\bottomrule
\end{tabular}
\end{table}

\begin{table}[ht]
\centering
\caption{Per-phase accuracy (\%): NL adversarial (120 runs).}
\label{tab:phases-adversarial}
\scriptsize
\begin{tabular}{@{}lrrrrrr@{}}
\toprule
Model & Disc. & C.Prec & C.Rec & Prof. & RDF & Feas. \\
\midrule
4.1-nano & 85 & 60 & 60 & 60 & 55 & 60 \\
4.1-mini & 95 & 85 & 90 & 90 & 90 & 90 \\
4.1      & 100 & 82 & 85 & 90 & 85 & 90 \\
5-nano   & 100 & 78 & 85 & 85 & 85 & 90 \\
5-mini   & 100 & 95 & 95 & 95 & 95 & 95 \\
5        & 100 & 75 & 75 & 95 & 75 & 95 \\
\bottomrule
\end{tabular}

\vspace{1mm}
{\scriptsize Disc.=Discovery, C.Prec/C.Rec=Constraint Precision/Recall, Prof.=Profile selection, RDF=valid RDF, Feas.=Feasibility pass.}
\end{table}

\section{RFSS Matching Algorithms}
\label{app:catalog}

Table~\ref{tab:catalog-perf} benchmarks the greedy heuristic against CP-SAT across three catalog scales (20 reps, 5 warmup). The greedy heuristic is 1.8$\times$ faster (Mann--Whitney $p < 10^{-3}$) with identical satisfaction rates; the cost gap is bounded and scale-dependent.

\begin{table}[ht]
\centering
\caption{Matching algorithm performance.}
\label{tab:catalog-perf}
\scriptsize
\begin{tabular}{@{}llrrrr@{}}
\toprule
Scale & Algo & Med (ms) & Mem (KB) & Cost & Sat\% \\
\midrule
Large  & Greedy & 78.8 & 230 & 1.0 & 25 \\
       & CP-SAT & 121.5 & 372 & 0.4 & 25 \\
Medium & Greedy & 25.2 & 93 & 1.6 & 26 \\
       & CP-SAT & 46.6 & 172 & 1.1 & 26 \\
Small  & Greedy & 0.5 & 7 & 1.9 & 29 \\
       & CP-SAT & 6.1 & 52 & 1.9 & 29 \\
\bottomrule
\end{tabular}
\end{table}

\section{Adversarial Breakdown and Context-Grounding Ablation}
\label{app:adversarial}
\label{app:ablation}

Table~\ref{tab:ablation} isolates the effect of injecting available-property metadata. Both configurations use identical adversarial prompts (120 runs) and self-healing loops; only the prompt context differs.

\begin{table}[ht]
\centering
\caption{Adversarial success (\%) without/with property injection.}
\label{tab:ablation}
\scriptsize
\begin{tabular}{@{}lrrr@{}}
\toprule
Model & Without & With & $\Delta$ \\
\midrule
GPT-4.1-nano  & 40 & 60 & +20 \\
GPT-4.1-mini  & 50 & 90 & +40 \\
GPT-4.1       & 75 & 90 & +15 \\
GPT-5-nano    & 70 & 90 & +20 \\
GPT-5-mini    & 50 & 95 & +45 \\
GPT-5         & 80 & 95 & +15 \\
\midrule
\textbf{Overall} & \textbf{61} & \textbf{87} & \textbf{+26} \\
\bottomrule
\end{tabular}

\vspace{1mm}
{\scriptsize Mini-tier models gain most (+40/+45); full-size models gain less (+15) as they already suppress unmodeled concepts.}
\end{table}

\begin{figure}[!ht]
\centering
\includegraphics[width=0.65\columnwidth]{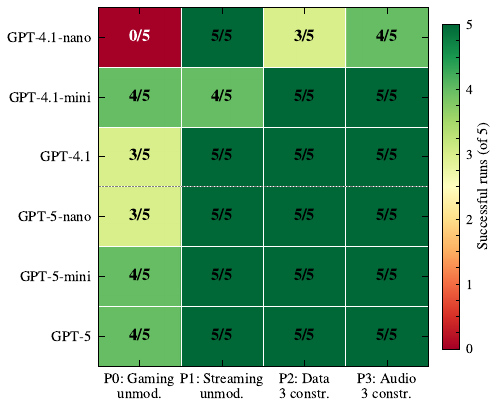}
\caption{Per-prompt adversarial success. ``unmod.'' = unmodeled metric. P2/P3 use only modeled metrics.}
\label{fig:adversarial-prompts}
\end{figure}

%% file: main.bbl
\begin{thebibliography}{20}
\providecommand{\natexlab}[1]{#1}
\providecommand{\url}[1]{\texttt{#1}}
\expandafter\ifx\csname urlstyle\endcsname\relax
  \providecommand{\doi}[1]{doi: #1}\else
  \providecommand{\doi}{doi: \begingroup \urlstyle{rm}\Url}\fi

\bibitem[Alcock et~al.(2025)Alcock, Anand, Rotsos, and Race]{alcock2025swift}
Alcock, P., Anand, R., Rotsos, C., and Race, N.
\newblock {SWIFT}: Semantic web intent framework for intent translation.
\newblock In \emph{2025 IEEE/IFIP Network Operations and Management Symposium
  (NOMS)}. IEEE, 2025.
\newblock \doi{10.1109/NOMS57970.2025.11073618}.

\bibitem[Boutouchent et~al.(2025)Boutouchent, Mekrache, Ksentini, Adhane,
  Fonseca, McNamara, Ramantas, Palena, Iordache, Lo~Cigno, Mostafa, Roy, and
  Verikoukis]{boutouchent20256gintense}
Boutouchent, A., Mekrache, A., Ksentini, A., Adhane, G., Fonseca, J., McNamara,
  J., Ramantas, K., Palena, M., Iordache, M., Lo~Cigno, R., Mostafa, S., Roy,
  S., and Verikoukis, C.
\newblock {6G-INTENSE}: Intent-driven native {AI} architecture supporting
  network-compute abstraction and sensing at the deep edge.
\newblock In \emph{Proceedings of the IEEE Vehicular Technology Conference
  (VTC-Spring)}. IEEE, 2025.
\newblock Preprint:
  \url{https://6g-intense.eu/wp-content/uploads/2025/01/Intense_VTM.pdf}.

\bibitem[Brenes et~al.(2025)Brenes, Piscione, Kolobov, and
  Ferrigno]{brenes2025unified}
Brenes, J., Piscione, P., Kolobov, M., and Ferrigno, G.
\newblock Unified intent-based management across standards: Architecture and
  prototype realizations.
\newblock In \emph{2025 IEEE Conference on Standards for Communications and
  Networking (CSCN)}. IEEE, 2025.
\newblock \doi{10.1109/CSCN67557.2025.11230699}.

\bibitem[Clemm et~al.(2022)Clemm, Ciavaglia, Granville, and Tantsura]{rfc9315}
Clemm, J., Ciavaglia, L., Granville, L., and Tantsura, J.
\newblock {RFC 9315: Intent-Based Networking - Concepts and Definitions}.
\newblock Rfc, IETF, 2022.
\newblock URL \url{https://www.rfc-editor.org/rfc/rfc9315}.

\bibitem[{ETSI}(2019)]{etsi-zsm-002}
{ETSI}.
\newblock {Zero-touch network and Service Management (ZSM); Reference
  Architecture}.
\newblock Etsi gs zsm 002, ETSI, 2019.
\newblock URL \url{V1.1.1}.

\bibitem[Ficzere et~al.(2025)Ficzere, Holl{\'o}si, and
  Varga]{ficzere2025beyond}
Ficzere, D., Holl{\'o}si, G., and Varga, P.
\newblock Beyond intent translation: Research gaps in the application of
  generative {AI} for intent-based networking.
\newblock In \emph{2025 IEEE Network Operations and Management Symposium (NOMS)
  Workshops --- Workshop on Generative AI in Network Management (GAIN)}. IEEE,
  2025.
\newblock \doi{10.1109/NOMS57970.2025.11073701}.

\bibitem[Guo et~al.(2025)Guo, Zhang, Wang, Wu, Yan, Sun, He, Qi, and
  Liao]{ibn-autonomous}
Guo, L., Zhang, L., Wang, J., Wu, J., Yan, Y., Sun, H., He, B., Qi, Q., and
  Liao, J.
\newblock Intent-based autonomous network framework guided by large language
  model.
\newblock \emph{IEEE Transactions on Automation Science and Engineering},
  22:\penalty0 22185--22197, 2025.
\newblock \doi{10.1109/TASE.2025.3610906}.

\bibitem[Jacobs et~al.(2025)Jacobs, Pfitscher, Ribeiro, Granville, Ferreira,
  Willinger, and Rao]{lumi-trust}
Jacobs, A.~S., Pfitscher, R.~J., Ribeiro, R.~H., Granville, L.~Z., Ferreira,
  R.~A., Willinger, W., and Rao, S.~G.
\newblock Establishing trust for using natural language for intent-based
  networking.
\newblock \emph{IEEE Transactions on Network and Service Management},
  22\penalty0 (5):\penalty0 4775--4787, 2025.
\newblock \doi{10.1109/TNSM.2025.3574626}.

\bibitem[Kapoor et~al.(2025)Kapoor, Gurbilek, Parra{-}Ullauri, Jangra, Khan,
  Duke, Hey, McHugh, and Corston{-}Petrie]{kapoor2025genai}
Kapoor, S., Gurbilek, G., Parra{-}Ullauri, J., Jangra, P.~K., Khan, T.~A.,
  Duke, A., Hey, A., McHugh, D., and Corston{-}Petrie, A.
\newblock {GenAI}-powered intent-based autonomous networks: Standardisation
  landscape and emerging industry trends.
\newblock \emph{TechRxiv}, 2025.
\newblock \doi{10.36227/techrxiv.175756256.63633045}.
\newblock preprint.

\bibitem[Leivadeas \& Falkner(2023)Leivadeas and Falkner]{ibnsurvey}
Leivadeas, A. and Falkner, M.
\newblock {A survey on Intent-Based Networking}.
\newblock \emph{IEEE Communications Surveys \& Tutorials}, 25\penalty0
  (1):\penalty0 625--655, 2023.

\bibitem[Martins et~al.(2026)Martins, Mokrushin, and
  Orlic]{martins2026tioshacl}
Martins, J., Mokrushin, L., and Orlic, M.
\newblock {TIO-SHACL}: Comprehensive {SHACL} validation for {TMF} intent
  ontologies, 2026.
\newblock URL \url{https://arxiv.org/abs/2604.27359}.
\newblock Open-source implementation:
  \url{https://github.com/EricssonResearch/tio-shacl}.

\bibitem[Mehmood et~al.(2023)Mehmood, Kralevska, and
  Palma]{mehmood2023knowledge}
Mehmood, K., Kralevska, K., and Palma, D.
\newblock Knowledge-based intent modeling for next generation cellular
  networks.
\newblock In \emph{IEEE International Mediterranean Conference on
  Communications and Networking (MeditCom)}. IEEE, 2023.

\bibitem[Mekrache et~al.(2024)Mekrache, Ksentini, and
  Verikoukis]{ibn-llm-centric}
Mekrache, A., Ksentini, A., and Verikoukis, C.
\newblock {Intent-based management of next-generation networks: an LLM-centric
  approach}.
\newblock \emph{IEEE Network}, 2024.

\bibitem[Mekrache et~al.(2025)Mekrache, Ksentini, and
  Verikoukis]{mekrache2025dmo}
Mekrache, A., Ksentini, A., and Verikoukis, C.
\newblock Dmo-gpt: An intent-driven framework for distributed 6g management and
  orchestration.
\newblock \emph{IEEE Communications Magazine}, 2025.

\bibitem[Niem\"{o}ller et~al.(2024)Niem\"{o}ller, M\"{u}ller, Maggiari, and
  Maghsoudlou]{ericsson-intent-evolution}
Niem\"{o}ller, J., M\"{u}ller, E., Maggiari, M., and Maghsoudlou, K.
\newblock {Evolving Service Management Toward Intent-Driven Autonomous
  Networks}.
\newblock \emph{Ericsson Technology Review}, 2024\penalty0 (3):\penalty0 2--7,
  2024.
\newblock \doi{10.23919/ETR.2024.10759715}.

\bibitem[Tageldien et~al.(2025)Tageldien, Selim, and
  Sboui]{tageldien2025llmibn}
Tageldien, M., Selim, B., and Sboui, L.
\newblock Large language models in intent-based networking: A comprehensive
  survey across the intent lifecycle.
\newblock In \emph{2025 International Telecommunications Conference
  (ITC-Egypt)}. IEEE, 2025.
\newblock \doi{10.1109/ITC-EGYPT66095.2025.11186656}.

\bibitem[{TM Forum}(2022)]{tmf-ig1253}
{TM Forum}.
\newblock {IG1253: Intent in Autonomous Networks}.
\newblock Introductory guide, TM Forum, 2022.
\newblock URL \url{https://www.tmforum.org/toolkits/intent/}.

\bibitem[{TM Forum}(2024{\natexlab{a}})]{tmf-tr290}
{TM Forum}.
\newblock {TR290: Intent Common Model}.
\newblock Technical report, TM Forum, 2024{\natexlab{a}}.
\newblock URL \url{https://www.tmforum.org/toolkits/intent/}.

\bibitem[{TM Forum}(2024{\natexlab{b}})]{tmf633}
{TM Forum}.
\newblock {TMF633: Service Catalog Management API v4.0}, 2024{\natexlab{b}}.
\newblock URL
  \url{https://www.tmforum.org/open-digital-architecture/open-apis/service-catalog-management-api-TMF633/v4.0}.

\bibitem[{TM Forum}(2024{\natexlab{c}})]{tmf641}
{TM Forum}.
\newblock {TMF641: Service Ordering Management API v5.0}, 2024{\natexlab{c}}.
\newblock URL
  \url{https://www.tmforum.org/open-digital-architecture/open-apis/service-ordering-management-api-TMF641/v5.0}.

\end{thebibliography}
